\documentclass[prl,aps,twocolumn]{revtex4}

\newcommand{\calL}{{\cal L}}

\usepackage{graphicx}
\usepackage{bm}

\begin{document}

\noindent
{\bf Comment on ``Confinement of Slave Particles in U(1) Gauge
Theories of Strongly Interacting Electrons''}

\medskip

\noindent
Masaki Oshikawa

\medskip

\noindent
Department of Physics, Tokyo Institute of Technology,
Oh-okayama, Meguro-ku, Tokyo 152-8551 JAPAN

\bigskip

In a recent Letter~\cite{Nayak1,Nayak2}, 
Nayak argued that the deconfinement can never happen
in the U(1) gauge theory approach, because
the ``holon'' charge $q_b$ and ``spinon'' charge $q_f$
can be assigned arbitrarily provided
$q_b - q_f = e$ ($-e$ is the electron charge).
This simple but important question seems never
answered appropriately~\cite{IM-comment,IMO}.
In this Comment, I shall examine the issue more carefully.

Besides the dynamical gauge field $a_{\mu}$,
I introduce (smooth) external static  gauge fields
$A^f_{\mu}$ and $A^b_{\mu}$ which
couple respectively to the spinon and holon fields.
To obtain the low-energy effective action,
high-energy parts of the dynamical fields are integrated out;
their low-energy parts
are kept fixed and are indistinguishable from the external fields.
As a consequence, under $A^f$ and $A^b$,
the effective action must be written in terms of 
$\tilde{A}^{\alpha}_{\mu} \equiv A^{\alpha}_{\mu} + a_{\mu}$.
Here I assume that the kinetic term of the gauge field is generated
in the low energy effective action, as it is argued in the U(1) gauge
theory approach.
The low-energy effective Lagrangian density is thus written as
\begin{eqnarray}
\calL
&=& \calL_f[\bar{f}_{\sigma},f_{\sigma},\tilde{A}_{\mu}^f] +
 \calL_b[\bar{b},b,\tilde{A}_{\mu}^b]
+ \frac{\lambda_f}{4} \tilde{F}_{\mu\nu}^f \tilde{F}_{\mu\nu}^f
\nonumber \\
&&
+ \frac{\lambda_m}{2} \tilde{F}_{\mu\nu}^f \tilde{F}_{\mu\nu}^b
+ \frac{\lambda_b}{4} \tilde{F}_{\mu\nu}^b \tilde{F}_{\mu\nu}^b
- a_{\mu} J^0_{\mu},
\end{eqnarray}
where $J^0_{\mu}$
represents the constant density of one particle per site.
Here
$\tilde{F}_{\mu\nu}^{\alpha} =
\partial_{\mu} \tilde{A}^{\alpha}_{\nu}
- \partial_{\nu} \tilde{A}^{\alpha}_{\mu}$ ($\alpha=f,b$)
and
$\lambda_{f,m,b}$ are coupling constants.
In the limit of zero external fields, we recover the usual
form with the gauge field coupling constant
$g^{-2} = \lambda_f + 2 \lambda_m + \lambda_b$.  

In the low-energy effective theory, one may consider
the currents $J^{f,b}$ defined by the functional derivative
of the action by the external fields $A^{f,b}$.
The equation of motion of the dynamical gauge field $a_{\mu}$
leads to the ``confinement constraint'' $J^f_{\mu}+J^b_{\mu}=J^0_{\mu}$.
However, these currents actually includes a contribution from
the gauge field:
$J^{f}_{\mu} = j^{f}_{\mu}
+ \lambda_f \partial_{\nu} \tilde{F}^f_{\mu \nu}
+ \lambda_m \partial_{\nu} \tilde{F}^b_{\mu \nu}$, 
where
$ j^f_{\mu} = {\partial \calL_f}/{\partial A^f_{\mu}}$
is the spinon current is in a standard definition.
The current $j^b$ is similarly defined.
The currents $j^{\alpha}$ do not obey
the confinement constraint.

To discuss the charge of the particles, I set
$A^f_{\mu} = q_f A_{\mu}, A^b_{\mu} = q_b A_{\mu}$,
where $A_{\mu}$ is the physical electromagnetic vector potential which
is regarded as an external field in this Comment.
Starting from the tentative assignment $q_f=0$ and $q_b=e$,
the redefinition of the gauge field as
$ a_{\mu}  \rightarrow a_{\mu} + c e A_{\mu}$
with an arbitrary parameter $c$
induces the change in the charges~\cite{Nayak1,Nayak2}
$q_f = c e$ and $q_b = (c+1) e$.
The physical current reads, after the redefinition,
$J^{em}_{\mu} = ce J^f_{\mu} + (1+c)e J^b_{\mu} - ce J^0_{\mu}$.
This apparently depends on the arbitrary parameter $c$.
However, the dependence on $c$ turns out to be
proportional to $J^f_{\mu} + J^b_{\mu} - J^0_{\mu}$, which vanishes
thanks to the equation of motion.
Thus the physical current is indeed independent of the
charge assignment, as it should be.

What is the charge of a slave particle, if it is deconfined?
For simplicity, below I take the limit of zero external field.
The total physical current is given by
$J^{em}_{\mu}
 = e J^b_{\mu} = e j^b_{\mu} +
 e (\lambda_m + \lambda_b) \partial_{\nu} f_{\mu \nu},$
independent of $c$.
Let us suppose that the spinon current $j^f_{\mu}$ is changed
by $\delta j^f_{\mu}$ while $j^b_{\mu}$ is kept fixed.
The change $\delta j^f_{\mu}$ affects the gauge field through
the equation of motion, and
the total change induced in the physical current reads
$\delta J^{em}_{\mu}
 =  - e \frac{\lambda_m + \lambda_b}{\lambda_f + 2 \lambda_m + \lambda_b}
\delta j^f_{\mu}$.
This means that, the effective charge of a deconfined spinon,
{\em including the contributions from
the gauge field},
is given by
$ Q_f = - e (\lambda_m + \lambda_b)/(\lambda_f + 2 \lambda_m + \lambda_b)$
which is independent of the arbitrary parameter $c$ of charge assignment.
A similar analysis gives the effective charge of
a deconfined holon $Q_b$, which satisfies
$Q_b - Q_f = e$.

To conclude,
the arbitrariness in the charge assignment observed by Nayak
does not contradict with the possibility of the deconfinement.
The effective charge carried by
a deconfined slave particle appears to be fractional,
which is independent of the arbitrary charge assignment
but is rather determined dynamically.
This fractional charge is apparently not quantized, unlike in the
fractional quantum Hall effect.
An estimate based on the one-loop calculation (see e.g.~\cite{NL})
shows~\cite{PAL}
that $Q_b$ and $Q_f$ are proportional to $(1-x)$ and $x$ respectively,
where $x$ is the doping concentration.
This implies that the ``holon'' indeed carries most (but not all!) of the
charge near half filling ($x \ll 1$).
More reliable non-perturbative calculations of
the effective charges $Q_{\alpha}$, along with the examinations
on whether the deconfinement really takes the place,
would be desired.

I thank Kazusumi Ino, Patrick A. Lee and Chetan Nayak for useful discussions.
This work is supported in part by Grant-in-Aid from MEXT of Japan.

\end{document}